\documentclass[aps,prl,reprint,groupedaddress,showpacs,showkeys]{revtex4-1}
\usepackage{watermark}
 \usepackage{amsmath}
 \usepackage{amsfonts}
 \usepackage{graphicx}
 \usepackage{epsfig}
 \usepackage{epstopdf}

 \usepackage{hyperref}
 \hypersetup{
    bookmarks=true,         
    unicode=false,          
    pdftoolbar=true,        
    pdfmenubar=true,        
    pdffitwindow=false,     
    pdfstartview={FitH},    
    pdftitle={Positron annihilation spectra and core-electron enhancement factors},   
    pdfauthor={D. G. Green},     
    pdfsubject={Positron_hlikeions},   
    pdfcreator={D. G. Green},   
    pdfproducer={Producer}, 
    pdfkeywords={keywords}, 
    pdfnewwindow=true,      
    colorlinks=true,       
    linkcolor=blue,          
    citecolor=blue,        
    filecolor=magenta,      
    urlcolor=blue           
}

\newcommand{\eps}{\varepsilon}
\newcommand{\au}{~a.u.}

\newcommand{\fig}[1]{Fig.~\ref{#1}}

\begin{document}
\title{$\gamma$-ray spectra and enhancement factors for positron annihilation with core electrons}
\author{D.~G. Green}
\email[Correspondences to:~]{dermot.green@balliol.oxon.org}
\altaffiliation{\newline Present address: {Joint Quantum Centre (JQC) Durham/ \hspace{-0.7ex}Newcastle}, Department of Chemistry, Durham University, South Road, Durham, DH1 3LE, UK.}
\author{G.~F. Gribakin}
\email{g.gribakin@qub.ac.uk}
\affiliation{Department of Applied Mathematics and Theoretical Physics, 
Queen's University Belfast, Belfast, BT7\,1NN, Northern Ireland, United Kingdom}
\date{\today}

\begin{abstract}
Many-body theory is developed to calculate the $\gamma$-spectra for positron annihilation with valence and core electrons in the noble gas atoms. A proper inclusion of correlation effects and core annihilation provides for an accurate description of the measured spectra [Iwata \textit{et al.}, Phys. Rev. Lett. {\bf 79}, 39 (1997)]. The theory enables us to calculate the
enhancement factors $\gamma_{nl}$, which describe the effect
of electron-positron correlations for annihilation on individual electron
orbitals $nl$. We find that the enhancement factors scale with the orbital
ionization energy $I_{nl}$ (in electron-volt), as $\gamma_{nl}=1+\sqrt{A/I_{nl}}+(B/I_{nl})^{\beta}$, where $A\approx 40$~eV,
$B\approx 24$~eV and $\beta\approx 2.3$.
\end{abstract}

\maketitle


\textit{Introduction.}---In this Letter we show that many-body theory provides an accurate description of the positron annihilation gamma spectra for noble gas atoms. Key to this is the ability of the theory to describe strong electron-positron correlations which enhance the annihilation beyond the independent-particle approximation (IPA) results. This enhancement is important not only for positron annihilation with weakly-bound valence electrons, but also for the core electrons, and the corresponding enhancement factors display a near-universal scaling with the electron orbital ionization energy.

Due to repulsion from the nuclei, low-energy positrons annihilate predominantly on the outermost (valence) electrons in atoms, molecules, and condensed matter systems. Small fractions of positrons can, however, tunnel through the repulsive potential and annihilate on the core electrons. Two-photon annihilation is the dominant mode in both cases. The corresponding Doppler-broadened $\gamma$-ray energy spectrum is centered on 511~keV, and is characteristic of the electron velocity distribution in the states involved. In particular, annihilation on the tightly-bound core electrons results in distinct features at higher-energy Doppler shifts in the spectrum \cite{PhysRevLett.38.241,PhysRevLett.79.39}. The core annihilation signal shows high elemental specificity \cite{PhysRevLett.77.2097}, which can be used to study vacancies in metals and identify defects in semiconductors \cite{PhysRevB.20.3566,PhysRevB.54.2397} (see \cite{RevModPhys.85.1583} and references therein). Positron annihilation on core electrons is also a key process in the surface-analytic positron-induced Auger-electron spectroscopy (PAES) \cite{PhysRevLett.61.2245, Ohdaira1997177,Weiss2007285,nepomuc3}, and the time-resolved PAES \cite{nepomucref}, which enables the study of dynamics of catalysis, corrosion, and surface alloying \cite{PhysRevLett.105.207401}. Coincident measurements of the annihilation $\gamma $-ray and Auger electrons allows one to determine the annihilation $\gamma $-ray spectra for individual core orbitals \cite{PhysRevLett.89.075503,PhysRevB.73.014114}.

Interpretation of the experiments relies heavily on theoretical input. 
For example, for PAES one needs to know the relative probabilities of positron annihilation with inner electrons of various atoms \cite{PhysRevB.41.3928}. 
However, the process of positron annihilation in many-electron systems is characterised by strong electron-positron correlations. These correlations affect both the positron wave function and the electron-positron annihilation vertex. They lead to dramatic enhancements of positron annihilation rates in heavier noble-gas atoms, compared with the single-particle (e.g., Hartree-Fock) approximation  (see \cite{PhysRevA.90.032712} and references therein), 
and have a significant effect on the shapes of the $\gamma$-ray spectra \cite{0953-4075-39-7-008,DGG_hlike,DGG_molgamma}. For atomic systems electron-positron correlations can be included systematically and accurately by using many-body theory methods \cite{PhysRevA.70.032720,PhysRevA.90.032712}. Many-body theory provided important initial insights into positron annihilation in metals by considering positrons in an electron gas \cite{PhysRev.129.1622,PhysRev.155.197}. These early works introduced the concept of \textit{enhancement factors}, which measure the increase of the electron density at the positron. Subsequently, density functional theories have been developed for condensed-matter systems \cite{PhysRevB.34.3820,RevModPhys.66.841}. They describe positron states and annihilation in real materials, often using parametrizations of the correlation energy and enhancement factors for the positron in electron gas from many-body theory \cite{Arponen1979343}. The enhancement factors are particularly large ($\sim $10) for the valence electrons, but they are also significant for the core electrons \cite{PhysRevB.20.883}. They can be used to improve the annihilation probabilities and $\gamma$-spectra calculated in IPA \cite{PhysRevLett.38.241,PhysRevB.41.3928}.  However, benchmarking common electron-density-dependent enhancement factors against accurate positron-atom calculations reveals their deficiencies \cite{PhysRevB.65.235103}. Their use also leads to spurious effects in the $\gamma$-ray spectra \cite{PhysRevB.54.2397}.

Positron interaction with noble-gas atoms has been studied thoroughly in experiment by measuring the scattering cross sections and annihilation rates.
This system is ideal for testing the ability of theoretical and computational approaches to account for electron-positron correlations. An extensive comparison with the available data attests the validity and accuracy of the many-body theory we have developed \cite{PhysRevA.90.032712}. An outstanding issue is the annihilation $\gamma $-spectra of Ar, Kr and Xe which were measured by the San Diego group \cite{PhysRevLett.79.39}, but till now have eluded theoretical description. In this work we extend the many-body theory approach to calculate the $\gamma$-ray spectra for positron annihilation with the valence and core electrons of the noble gas atoms. We show that by properly accounting for the correlations in both the core and valence annihilation, the theory yields excellent agreement with experiment, including the range of large Doppler shifts where the core contribution dominates. The many-body theory also allows us to quantify the effect of correlations on the annihilation vertex and extract the ``exact'' enhancement factors $\gamma_{nl}$ for individual valence and core electron orbitals $nl$ \footnote{The theory also allows one to extract momentum-dependent enhancement factors \cite{DGG_molgamma,DGG_hlike}}.

\begin{figure}[ht!]
\includegraphics[width=0.475\textwidth]{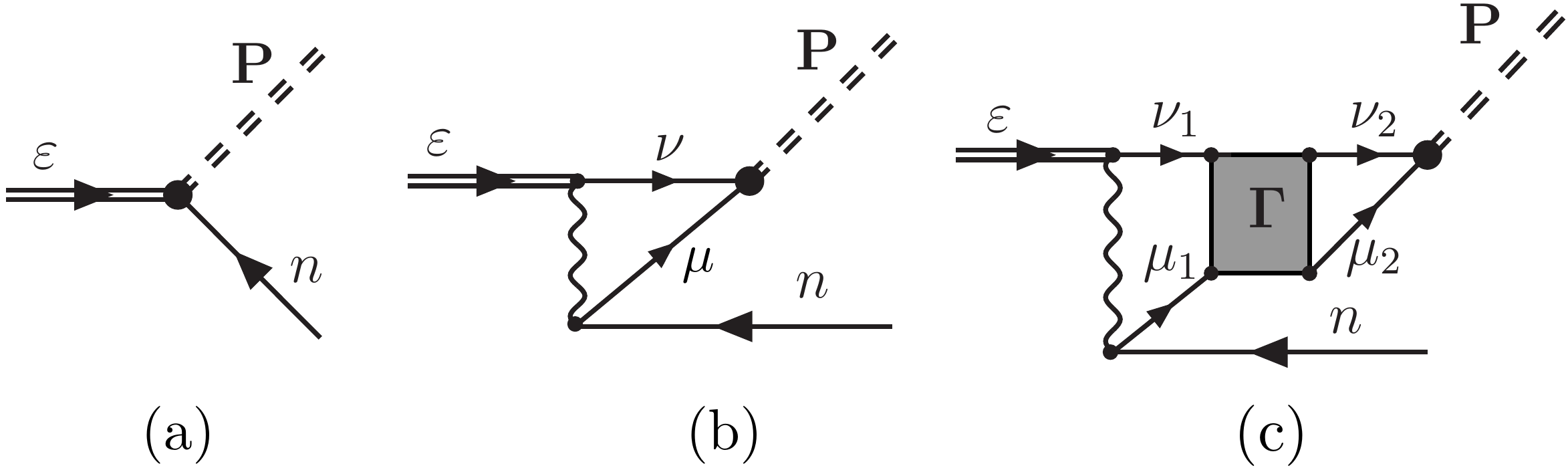}
\caption{Amplitude of positron annihilation with an electron in state $n$:
(a) zeroth-order vertex, (b) first-order, and (c) `$\Gamma$-block' corrections. Double lines labelled $\varepsilon$ represent the incident positron wave function; single lines labelled $\nu$ (${\mu}$) represent positron (excited electron) states, which are summed over; lines labelled $n$ represent holes in the atomic ground state; wavy lines represent the electron-positron Coulomb interactions, and double-dashed lines represent the two $\gamma$-ray photons. The $\Gamma$-block is the sum of an infinite series of electron-positron ladder diagrams \cite{PhysRevA.90.032712,PhysRevA.70.032720}.\label{fig:annamp}}
\end{figure}

\textit{Theory.}---In the dominant process, a positron annihilates with an electron in state $n$ to form two $\gamma$-ray photons of total momentum ${\bf P}$ \cite{QED}. In the centre-of-mass frame (${\bf P}=0$), the two $\gamma$-rays have equal energies $mc^2=511$\,keV (neglecting the initial positron and electron energies $\eps $ and $\eps_n$). In the laboratory frame, however, the photon energies are Doppler shifted by $\epsilon\leq Pc/2$ (typically, a few keV). The corresponding $\gamma$-spectrum is
\begin{eqnarray}\label{eqn:gammaspectra}
w_n(\epsilon)
=\frac{1}{c} \int _{2|\epsilon|/c}^{\infty} \int_{\Omega_{\bf P}} |A_{n\varepsilon}({\bf P})|^2 \frac{ d\Omega_{\bf P}}{(2\pi)^3} PdP,
\end{eqnarray}
where $A_{n\varepsilon}({\bf P})$ is the annihilation amplitude  \cite{0953-4075-39-7-008}. It is given diagrammatically in \fig{fig:annamp} (see \cite{0953-4075-39-7-008,DGG_hlike,PhysRevA.90.032712,DGG_corePRA,DGG_thesis} for details).
Specifically, shown are the dominant contributions to the annihilation vertex: the zeroth-order vertex, which represents the IPA amplitude, the first-order, and higher-order (`$\Gamma$-block') corrections, which account for the attractive electron-positron interaction at short range and enhance the annihilation rate. In practice, one usually calculates the spectrum for \textit{all} electrons in a given atomic orbital $nl$. The total spectrum, $w(\epsilon)=\sum_{nl}w_{nl}(\epsilon)$, which is probed in experiment, retains distinct features of the contributions of individual valence and core orbitals.

The fully-correlated incident positron \emph{quasiparticle} wave function $\psi_{\varepsilon}$ is calculated from the Dyson equation
$\bigl(H_0+\hat{\Sigma}_{\varepsilon}\bigr)\psi_{\varepsilon}=\varepsilon\psi_{\varepsilon}$, where $H_0$ is the Hamiltonian of the positron in the field of the Hartree-Fock (HF) ground state atom, and $\hat{\Sigma}_{\varepsilon}$ is the positron self-energy operator which plays the role of a nonlocal, energy-dependent positron-atom correlation potential. This potential accounts for polarization of the atom by the positron and for virtual positronium formation (represented by the $\Gamma $-block), both of which contribute to the positron-atom attraction (see \cite{PhysRevA.90.032712,PhysRevA.70.032720} for details).

The positron annihilation rate in a gas is usually parameterized by 
the dimensionless effective number of electrons, $Z_{\rm eff}$ \footnote{$Z_{\rm eff}$ is the ratio of the positron annihilation rate in an atomic or molecular gas to the basic Dirac annihilation rate in the electron gas of the same number density.}. For an orbital $nl$, it is given by
\begin{equation}\label{eqn:zeffspec}
Z_{{\rm eff},nl}=\int_{-\infty}^{\infty}  {w}_{nl}(\epsilon)\,d\epsilon.
\end{equation}
In general, $Z_{{\rm eff},nl}$ for valence electrons is greater than the number of electrons in the subshell, owing to the positron-atom attraction and enhancement due to the electron-positron short-range correlations.

The positron self-energy diagrams and the annihilation amplitude involve summations over intermediate excited electron and positron continuum states.  We calculate them numerically by employing B-spline basis sets.  Here, we use a basis of 40 splines of order 6, and a spherical box of radius 30 a.u. The maximum angular momentum of the intermediate states is $l_{\rm max}$=15. For this basis the sums over the energies converge rapidly, and we perform extrapolation to $l_{\rm max}\to\infty$ as in \cite{0953-4075-39-7-008}. Full details of the numerical implementation are given in \cite{PhysRevA.90.032712,DGG_thesis,DGG_corePRA}.

\textit{Results.}---The annihilation $\gamma$-ray spectra for Ar, Kr and Xe were measured at low gas pressures with room-temperature positrons confined in a Penning-Malmberg trap \cite{PhysRevLett.79.39}. That work also showed that the IPA [\fig{fig:annamp}~(a)] fails to describe the spectra accurately. It overestimates both the full width at half maximum (FWHM) of the spectra by 10--15\%, and the fraction of core annihilation, as seen from an excessive spectral weight at large Doppler shifts. Ref.~\cite{0953-4075-39-7-008} showed that the first-order correction [\fig{fig:annamp}~(b)] led to a narrowing of the spectrum, but was insufficient to describe the measured spectra.

\begin{figure}[t!]
\begin{center}
\includegraphics*[width=0.4\textwidth]{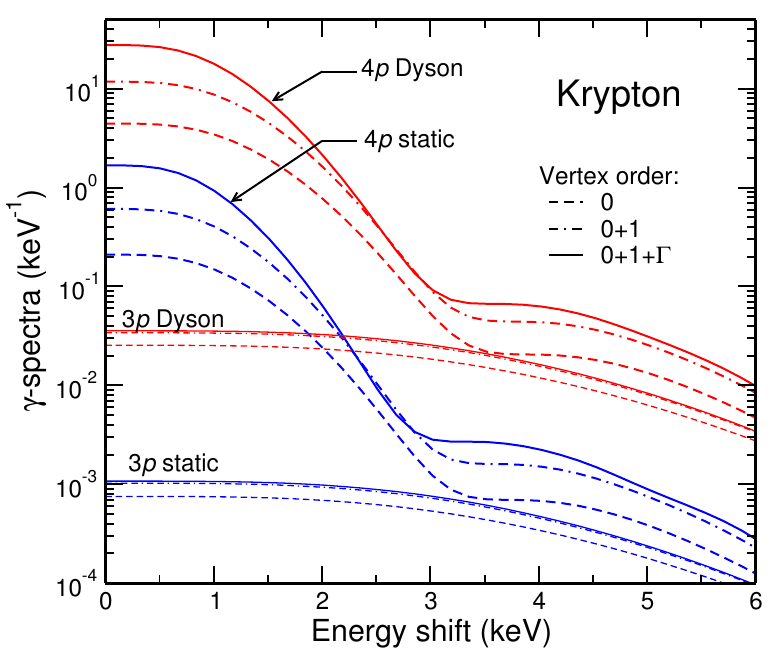}
\caption{
Annihilation $\gamma$-ray spectra for the $4p$ valence and $3p$ core electron orbitals in Kr, calculated using the positron wave function in the static field of the HF atom, and with the account of the correlation potential $\hat \Sigma _\eps$ (Dyson), and with various approximations for the annihilation vertex [\fig{fig:annamp}]. Dashed curves are for the zeroth-order vertex (``0'', IPA); chain curves include the first-order correction (``$0+1$''); solid curves show the results for the full vertex (``$0+1+\Gamma $'').
\label{fig:Kr4p3p}
}
\end{center}
\end{figure}

The full calculation presented in this work highlights the importance of higher-order corrections [\fig{fig:annamp}~(c)], especially for the valence electrons. The many-body theory also shows that the self-energy correlations that affect the positron wave function (double line in \fig{fig:annamp}) and the correlation corrections to the annihilation vertex [diagrams (b) and (c)]
have strikingly different effects on the spectra. As an example, \fig{fig:Kr4p3p} presents the $\gamma$-ray spectra for the outer valence $4p$ orbital and a core $3p$ orbital in Kr \footnote{All calculations are done for the $s$-wave incident positron with room-temperature momentum $k=0.04$\au}. It shows that the vertex corrections enhance the annihilation signal
by almost an order of magnitude for the valence electrons and by about 50\% for the core orbital. The role of the higher-order corrections [\fig{fig:annamp}~(c)] is much more prominent for the valence electrons.
Vertex corrections also lead to a significant narrowing of the spectrum for the valence electrons. Physically, this is related to the fact that in the vertex correction diagrams the positron annihilates with an excited electron, whose wave function is more diffuse than that of the core hole.

In contrast, the main result of improving the positron wave function (i.e., using the Dyson orbital rather than the static HF wave function) is a uniform increase in the annihilation signal. This increase is due to the build-up of the positron density in the vicinity of the atom caused by the positron-atom attraction. The magnitude of this effect is similar for the valence and core electrons. However, in contrast with the vertex corrections, it is sensitive to the atomic environment and the positron energy. In particular, low-energy annihilation in heavier noble-gas atoms is strongly enhanced by positron virtual states \cite{PhysRevA.90.032712}.

\begin{figure}[t!]
\begin{center}
\includegraphics*[width=0.35\textwidth]{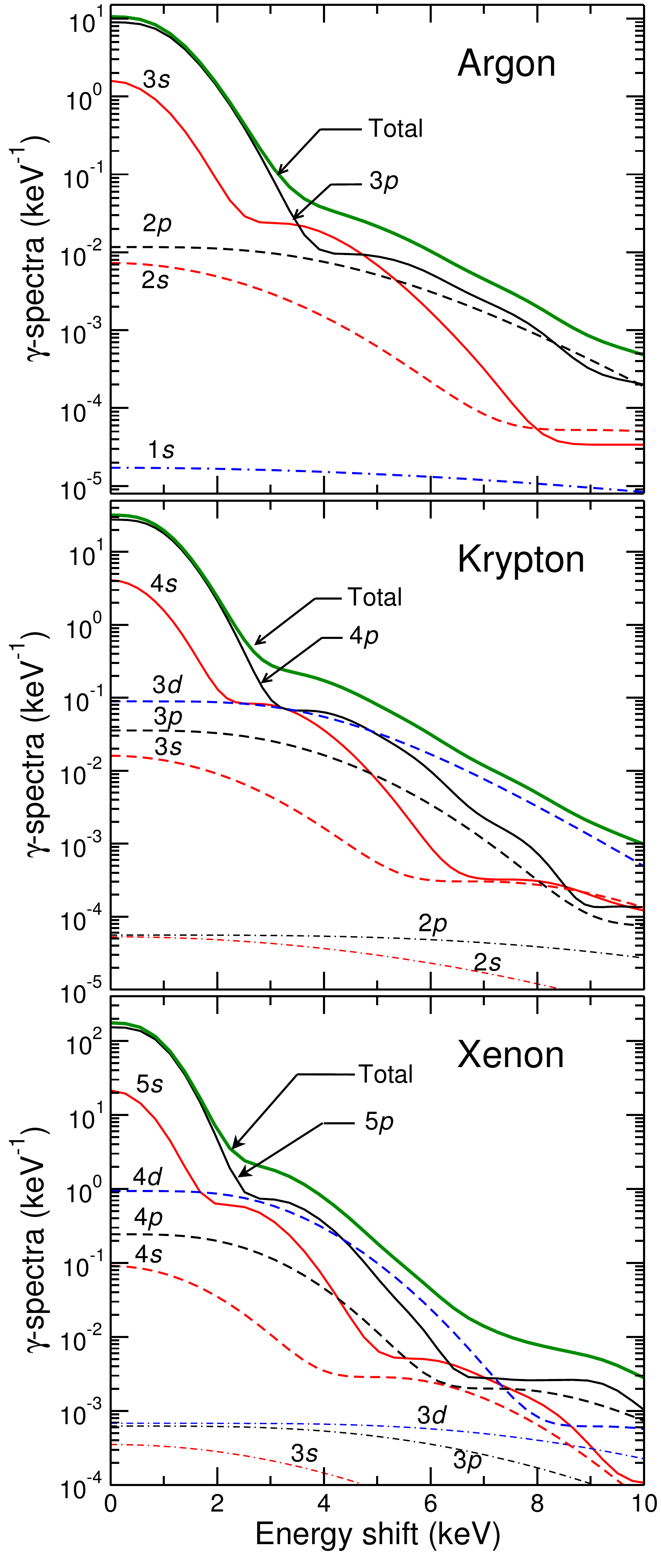}
\caption{Calculated $\gamma$-spectra for positron annihilation on individual subshells $nl$ in Ar, Kr and Xe: valence $ns$, $np$, (solid black and red lines); core $(n-1)s$, $(n-1)p$, and $(n-1)d$ (dashed lines); inner core $(n-2)s$, $(n-2)p$, and $(n-2)d$ (dash-dash-dotted lines); and total spectra (thick solid green line). All spectra are obtained using the full annihilation vertex (\fig{fig:annamp}) and Dyson positron wave function.}
\label{fig:spectradetailed}
\end{center}
\end{figure}

Figure \ref{fig:spectradetailed} shows the $\gamma$-spectra for positron annihilation on individual subshells of Ar, Kr and Xe, calculated with the full amplitude (\fig{fig:annamp}) using the Dyson $s$-wave positron state of thermal momentum $k=0.04$~a.u. The narrowly peaked valence spectra dominate the total spectra at low Doppler shifts. Compared with the valence electrons, the tightly-bound core electrons have greater velocities and produce broader $\gamma$-ray spectra. Note also that most individual spectra include multiple `shoulders'. The number of these is determined by the number of nodes $n-l$ in the electron radial wave function, as each `lobe' of the bound state wave function produces a characteristic contribution to the annihilation momentum density. In this way the spectra of the valence orbitals contain high-momentum components characteristic of the core orbitals to which they are orthogonal. Overall, the total $\gamma$-spectra retain the characteristics of the valence and core contributions.

\begin{figure}[t!]
\begin{center}
\includegraphics*[width=0.35\textwidth]{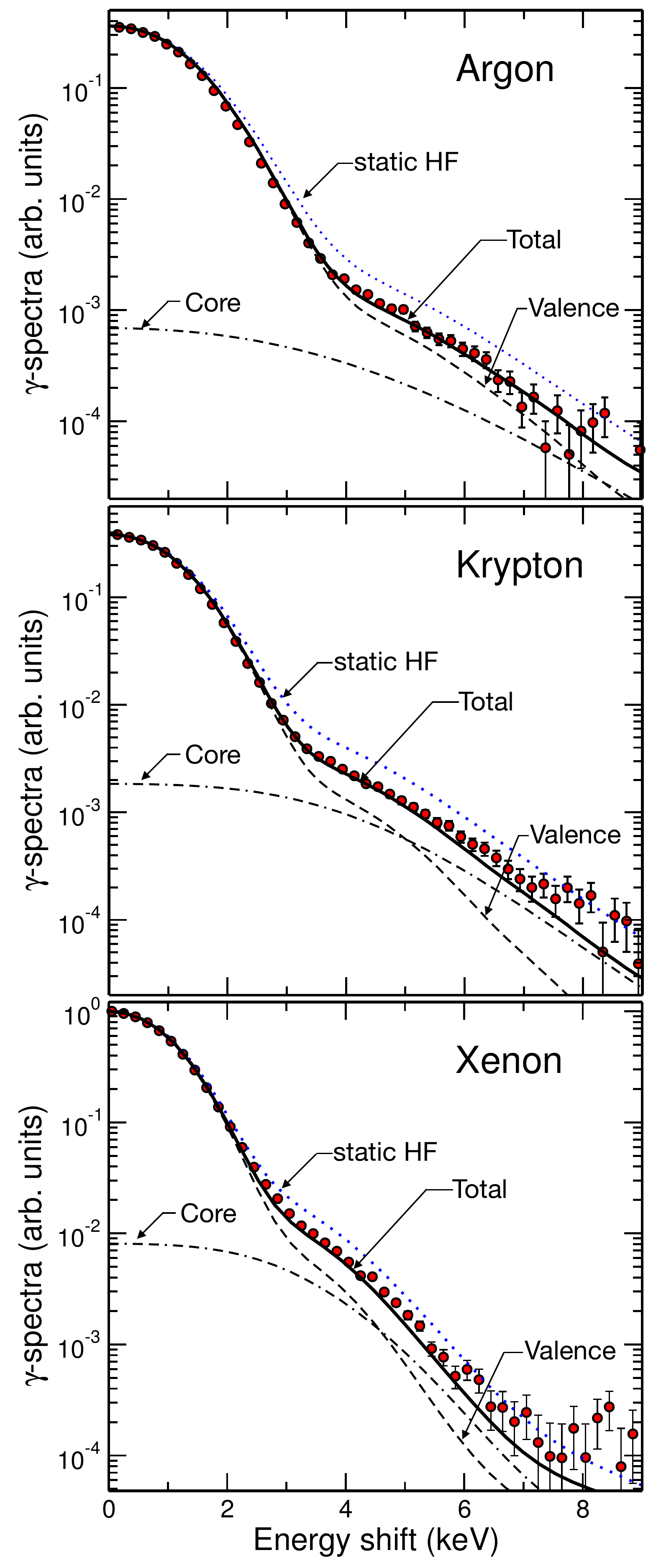}
\caption{$\gamma$-spectra for positron annihilation in Ar, Kr and Xe. Experiment: red circles. Theory (convolved with the detector resolution function): valence (dashed line); core (dash-dash-dotted line); and total (solid line), calculated with the full annihilation vertex and Dyson positron wave function. Also shown is the static HF (IPA) calculation of \cite{PhysRevLett.79.39} (blue dots).}
\label{fig:spectraexpt}
\end{center}
\end{figure}

Figure \ref{fig:spectraexpt} shows the calculated total spectra convolved with the detector resolution function and normalized to the experimental data at zero Doppler shifts \cite{PhysRevLett.79.39}. 
For each atom the valence component underestimates the experimental spectrum in the high-energy `wings', while the inclusion of the core brings the theoretical spectra into close agreement with experiment \footnote{In Kr and Xe the theoretical spectrum slightly underestimates the measurements at large Doppler shifts. One possible reason for this discrepancy is the neglect of relativistic effects on the electron wave functions \cite{PhysRevLett.52.1116}. However, a recent work \cite{PhysRevA.90.042702} which employs model potentials to describe positron-atom interactions, shows that the relativistic effect on the $\gamma $-spectra is small, e.g., increasing the FWHM in Xe by only 1.4\%}. This agreement supports the accuracy of the fractions of core annihilation derived from our many-body theory calculations: 0.55\% in Ar, 1.53\% in Kr, and 2.23\% in Xe \footnote{We estimate the uncertainty in these numbers to be about 5\%, comparable to effect of nonladder 3rd-order diagrams on the valence annihilation rates \cite{PhysRevA.90.032712}.}.

Note that the IPA $\gamma $-spectra obtained with the positron states in the static atomic field (dotted lines in \fig{fig:spectradetailed}) are significantly broader than the experiment. Such calculation also overestimates the fraction of core annihilation by a factor of two. However, when this fraction is used as a free parameter to fit the experimental data \cite{PhysRevLett.79.39}, the core annihilation fractions for Kr and Xe (1.3\% and 2.4\%, respectively) are close to the above \textit{ab initio} values.

\begin{figure}[t!]
\includegraphics*[width=0.4\textwidth]{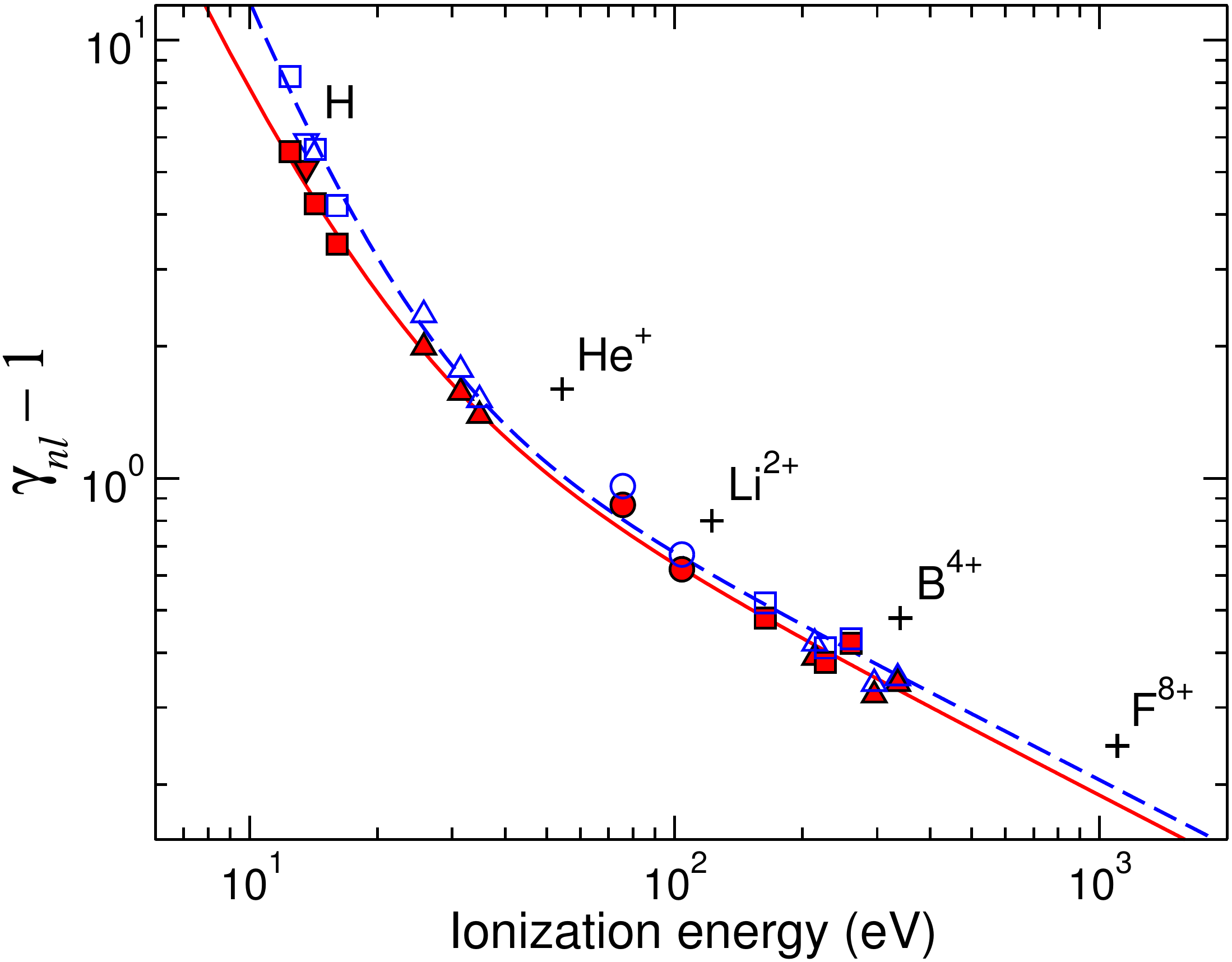}
\caption{Enhancement factors (\ref{eqn:enhfacszeff}) calculated using static HF (open symbols) and Dyson (solid symbols) positron states, for the $ns$ (triangles), $np$ (squares) and $nd$ (circles) valence and core orbitals in Ar, Kr, and Xe; $1s$ orbitals hydrogen (upside-down triangles)\cite{PhysRevA.70.032720}; and hydrogen-like ions (plus signs) \cite{DGG_hlike}. Dashed line is the fit (\ref{eqn:enhancefit}) of $\gamma _{nl}$ for atoms obtained using the static HF positron wave function ($A=42.0$~eV, $B=24.9$~eV, $\beta=2.54$), and the solid line is that for the Dyson positron wave function ($A=35.7$~eV, $B= 22.7$~eV, $\beta=2.15$).}
\label{fig:enhfacs}
\end{figure}

\textit{Enhancement factors.}---The many-body theory developed here and in Ref.~\cite{PhysRevA.90.032712} allows us to calculate the enhancement factors due to the correlation corrections to the annihilation vertex (\fig{fig:annamp}). These factors can be determined from the ratio of the annihilation rates $Z_{\rm eff}$ obtained with the full vertex to that of the zeroth-order (IPA), for each electron orbital $nl$:
\begin{equation}\label{eqn:enhfacszeff}
\gamma_{nl}=\frac{Z_{{\rm eff},nl}^{(0+1+\Gamma )}}
{Z_{{\rm eff},nl}^{(0)}}.
\end{equation}
Figure \ref{fig:enhfacs} shows the enhancement factors $\gamma_{nl}$ for the core and valence orbitals of Ar, Kr and Xe, for both static HF and Dyson incident positron states. Also shown are values of $\gamma _{1s}$ for hydrogen and hydrogen-like ions, obtained using the many-body theory approach \cite{PhysRevA.70.032720,DGG_hlike}.

The values of $\gamma_{nl}$ obtained with the positron wave function in the static atomic field are slightly larger that those found using the fully correlated Dyson wave functions (although this effect is negligible for the positive ions). This difference aside, \fig{fig:enhfacs} displays a near-universal scaling of the enhancement factors for the neutral atoms with the orbital ionization energy $I_{nl}$. This scaling can be parametrized by the formula
\begin{equation}\label{eqn:enhancefit}
\gamma_{nl}=1+\sqrt{A/I_{nl}}+ \left(B/I_{nl}\right)^\beta ,
\end{equation}
where $A$, $B$ and $\beta$ are constants found by fitting the numerical data.
The second term on the right-hand side of (\ref{eqn:enhancefit}) describes the effect of the first-order correction, \fig{fig:annamp}~(b). Its scaling with $I_{nl}$ is motivated by the $1/Z$ scaling of the enhancement factors in hydrogen-like ions \cite{DGG_hlike}. The third term is phenomenological; it accounts for the higher-order corrections which are important for the valence electrons (cf. \fig{fig:Kr4p3p}).

\textit{Summary.---}Many-body theory has been used to calculate the contribution of individual subshells to the $\gamma$-spectra of positron annihilation in noble gases. Inclusion of core annihilation gives results in excellent agreement with experiment and yields accurate core annihilation probabilities. `Exact' vertex enhancement factors obtained from the calculations have been found to follow a simple scaling with the electron ionization energy. We suggest that this result can be used to improve simple IPA calculations of core annihilation on atoms across the periodic table and in condensed matter.

\begin{acknowledgments}
We thank C.~M.~Surko for valuable discussions. DGG is grateful to the Institute for Theoretical Atomic, Molecular and Optical Physics at the Harvard-Smithsonian Centre for Astrophysics, where he carried out part of this work, and thanks H.~R.~Sadeghpour and colleagues for their generous hospitality.
\end{acknowledgments}


%

\end{document}